\documentclass[conference]{IEEEtran}
\IEEEoverridecommandlockouts
\usepackage{cite}
\usepackage{amsmath,amssymb,amsfonts} 
\usepackage{textcomp}
\usepackage{xcolor}
\usepackage{tikz}
\usepackage{amsmath}
\usepackage{amssymb,amsfonts,amsmath}
\usepackage{textcomp}
\usepackage{csquotes}
\usepackage{booktabs}
\usepackage{color}
\usepackage{epstopdf} 
\usepackage[capitalize]{cleveref}
\usepackage{url}
\usepackage{xspace}
\usepackage{times}
\usepackage{epsfig}
\usepackage{blindtext}
\usepackage{amsmath}
\usepackage{amssymb}
\usepackage{algpseudocode}
\usepackage{ulem}
\usepackage{mathtools}
\usepackage{ifthen}
\usepackage{cite} 
\usepackage[boxruled]{algorithm2e} 
\usepackage{graphicx}
\usepackage[capitalize]{cleveref}
\usepackage{float}
\usepackage{caption}
\usepackage{subcaption}
\usepackage{amsthm}
\usepackage{setspace}
\usepackage{enumitem}
\theoremstyle{definition}

\newcommand{\sys}{Baxos\xspace}
\newcommand{\synod}{Synod Paxos\xspace}

\usetikzlibrary{calc}
\usetikzlibrary{decorations}
\usetikzlibrary{positioning}
\usetikzlibrary{shapes}
\usetikzlibrary{calligraphy}
\def\BibTeX{{\rm B\kern-.05em{\sc i\kern-.025em b}\kern-.08em
    T\kern-.1667em\lower.7ex\hbox{E}\kern-.125emX}}
\begin{document}

\title{Baxos: Backing off for Robust and Efficient Consensus}

\author{
  \makebox[.5\linewidth]{Pasindu Tennage}\\Ecole Polytechnique Fédérale de Lausanne (EPFL)\\
  \and \makebox[.5\linewidth]{Cristina Basescu}\\Ecole Polytechnique Fédérale de Lausanne (EPFL)\\
  \and \makebox[.5\linewidth]{Eleftherios Kokoris Kogias}\\IST Austria and Mysten Labs\\
  \and \makebox[.5\linewidth]{Ewa Syta}\\Trinity College\\
  \and \makebox[.5\linewidth]{Philipp Jovanovic}\\University College London\\
  \and \makebox[.5\linewidth]{Bryan Ford}\\Ecole Polytechnique Fédérale de Lausanne (EPFL)\\

}

\maketitle
\thispagestyle{plain}
\pagestyle{plain}

\begin{abstract}

Leader-based consensus algorithms are vulnerable to liveness and performance downgrade attacks. We explore the possibility of replacing leader election in Multi-Paxos with \textit{random exponential backoff (REB)}, a simpler approach that requires minimum modifications to the two phase Synod Paxos and achieves better resiliency under attacks.

We propose \sys, a new resilient consensus protocol that leverages a random exponential backoff scheme as a replacement for leader election in consensus algorithms. Our backoff scheme addresses the common challenges of random exponential backoff such as scalability and robustness to changing wide area latency. We extensively evaluate \sys to illustrate its performance and robustness against two liveness and performance downgrade attacks using an implementation running on Amazon EC2 in a wide area network and a combination of a  micro benchmark and YCSB-A workload on Redis. Our results show that \sys offers more robustness to liveness and performance downgrade attacks than leader-based consensus protocols.
\sys outperforms Multi-Paxos and Raft up to 128\% in throughput under liveness and performance downgrade attacks under worst case contention scenarios where each replica proposes requests concurrently while only incurring a 32\% reduction on the maximum throughput in the synchronous attack-free scenario.
    
\end{abstract}

\section{Introduction}\label{sec:intro}

Consensus is a widely used abstraction to ensure strong consistency in distributed systems. When run in multiple instances, which is also known as state machine replication (SMR)~\cite{schneider1990implementing}, consensus enables a set of replicas to agree on a single history of operations. Popular systems that use consensus include Chubby~\cite{burrows2006chubby}, ZooKeeper \cite{hunt2010zookeeper} and Boxwood~\cite{maccormick2004boxwood}.

For performance reasons, most deployed consensus protocols use a leader which serves client requests and inter-replica messages~\cite{chandra2007paxos,ongaro2014search,junqueira2011zab}. In particular, the leader is tasked with handling contention and providing lock-free termination~\cite{lamport2001paxos} which works well in synchronous and attack-free network settings. However, under more adversarial network conditions, this approach becomes problematic~\cite{nikolaou2015turtle,spiegelman2019ace}. When the network is volatile, e.g., changing link delays and bandwidth, leader-based approaches fail to deliver good performance due to leader timeouts and subsequent leader election mechanisms which impact the overall system availability. In the worst case a service can even freeze completely, which is exactly what happened in a recent outage at Cloudflare~\cite{cloudflare}. This downside becomes particularly problematic when a system is under a distributed denial of service (DDoS) attack. Previous research has shown that even a weak adversary, who can attack only a single replica at a time, can halt the availability of a leader-based consensus algorithm~\cite{nikolaou2016moving,spiegelman2019ace}. With DDoS attacks becoming more prevalent~\cite{douligeris2004ddos}, leader-based consensus algorithms pose a significant risk to the availability of Internet applications.

Previously proposed consensus algorithms that achieve lock-free termination without using a leader node include multi-leader-based protocols~\cite{barcelona2008mencius,biely2012s,zhao2018sdpaxos,charapko2021pigpaxos}, sharding based protocols~\cite{marandi2012multi,ailijiang2017wpaxos,lockerman2018fuzzylog,corbett2013spanner}, application dependent protocols that exploit request dependencies~\cite{lamport2005generalized,park2019exploiting,moraru2013there} and asynchronous algorithms~\cite{aspnes2003randomized}. However, these approaches have liveness and performance vulnerabilities. Most multi-leader-based algorithms delegate message propagation to other replicas but still rely on a leader to order requests~\cite{biely2012s,zhao2018sdpaxos,charapko2021pigpaxos}, remaining susceptible to attacks on the leader node.
Algorithms that exploit request dependencies are vulnerable to DDoS attacks that issue concurrent dependent requests~\cite{moraru2013there,tollman2021epaxos}. 
DDoS attacks against the top level shards in sharding-based consensus algorithms 
~\cite{marandi2012multi,ailijiang2017wpaxos,lockerman2018fuzzylog,corbett2013spanner} can make the entire system unavailable. Finally, fully asynchronous algorithms are generally complex, rarely implemented, and usually do not perform as well as Multi-Paxos in practice.

We observe that even after two decades of leaderless consensus protocol research, the majority of the deployed consensus algorithms still use leader-based protocols such as Multi-Paxos or Raft. This situation has led us to investigate the minimal modification required to transform a consensus algorithm such as Synod-Paxos~\cite{lamport2001paxos} to a consensus algorithm that is robust against liveness and performance downgrade attacks while preserving good performance in an attack-free scenario.
In turn, we explore the possibility of utilizing random exponential backoff (REB)~\cite{bender2016scale} in the context of consesus,
due to its robustness, efficient contention handling, and power efficiency guarantees. As a result, we propose \sys, a bare minimal modification of Synod-Paxos~\cite{lamport2001paxos} that is robust and highly available under liveness and performance downgrade attacks. 

\sys employs the same two-phase protocol core as Synod Paxos, but in contrast to Multi-Paxos, it uses REB instead of leader election to achieve lock-free termination. In \sys, every node can propose values and, when concurrent proposals collide, they \textit{back off} to avoid further collisions, an approach similar to CSMA in LANs~\cite{ziouva2002csma}. Replacing leader election with random exponential backoff is not trivial, however, due to its potential side effects such as (1) the \textit{capture effect}, where a single node can have an unfair share of a shared resource as well as (2) the impact on resilience to changing network delays and (3) scalability. \sys leverages a REB protocol that scales up to nine replicas while remaining resilient to changing network delays and minimizing the capture effect.

\sys is the first attempt to prototype REB-based Paxos and to systematically explore its properties.
To evaluate the properties of \sys, we compare \sys against Multi-Paxos~\cite{lamport2001paxos} and Raft~\cite{ongaro2014search}
. We first analyze the performance of \sys under \textit{delayed view change attacks}, a class of targeted performance downgrade attacks in the wide area, and show that \sys, in such a situation, significantly outperforms Multi-Paxos and Raft by up to 128\% in throughput. Then, we explore the performance overhead of \sys under attack free synchronous network scenarios in the wide-area, and show that it achieves a throughput of 17,500 requests per second in contrast to the 28,000 requests per second saturation throughput of Multi-Paxos and Raft. Third, we analyze the uniformity of bandwidth utilization and show that \sys achieves a more uniform resource utilization across a set of consensus replicas than Multi-Paxos and Raft. Finally we show that Baxos can scale up to nine nodes in the wide-area.

To summarize, this paper makes the following contributions: 
\begin{enumerate}[noitemsep,topsep=0pt,parsep=0pt,partopsep=0pt]
	\item We explore the use of REB as a replacement for the leader election in consensus algorithms.
	\item We design and systematically develop a consensus algorithm by combining Synod Paxos and REB.
	\item We provide an experimental analysis of \sys performance under both adversarial and normal-case network conditions as well as a bandwidth resource utilization analysis.
\end{enumerate}
\section{Background}\label{sec:back}
This section provides an overview of the consensus problem, consensus algorithms, including leader-based protocols and its performance vulnerabilities, as well as the random exponential backoff mechanism we use as a building block of \sys.

\subsection{Consensus}

Consensus is an abstraction used to reach an agreement among a set of replicas. A consensus protocol allows each node to propose a value, agree upon one of the proposed values and to report it to all live replicas. 

A correct consensus algorithm satisfies four main properties~\cite{cachin2011introduction}: 
(1) \textit{validity}, a decided value should be previously proposed by a node;
(2) \textit{termination}, every correct process eventually decides some value; 
(3) \textit{integrity}, no process decides twice; and 
(4) \textit{agreement}, no two correct processes decide differently. 

We focus on non-Byzantine consensus, where nodes are cooperative (non-malicious), although the network can be adversarial. 
The FLP theorem~\cite{fischer1985impossibility} states that consensus is impossible in the asynchronous network setting even in the presence of a single node failure. Practical consensus algorithms alleviate the FLP impossibility result using a partial synchrony assumption or randomization.

State machine replication (SMR) is a use-case of consensus, where nodes run multiple instances of consensus to agree on a series of values~\cite{schneider1990implementing}. Consensus and SMR are generally considered to be equivalent, but from a theoretical perspective SMR is more expensive than consensus in terms of complexity and instructions~\cite{antoniadis2018state}.

\subsection{Leader-Based Consensus}

Multi-Paxos~\cite{lamport2001paxos} and Raft~\cite{ongaro2014search} are the most widely deployed consensus algorithms that rely on partial synchrony to alleviate the FLP impossibility result. Multi-Paxos builds on top of Synod Paxos by having a leader replica that handles client requests. A  replica runs the Prepare-Promise phase for a batch of consensus instances in the leader election phase and becomes the leader. Then, each client request is committed in the Propose-Accept phase in a single round trip. Raft builds on top of view-stamp replication~\cite{oki1988viewstamped}. 
When the leader is stable, Raft achieves a single round trip time consensus. When the leader fails, Raft uses a leader election algorithm to elect a new leader. On a high level, both Multi-Paxos and Raft solve the consensus problem in a similar method, differing only in the way a new leader is elected~\cite{howard2020paxos}. 

\subsection{Performance Vulnerabilities}~\label{ddos_background}
Consensus protocols are often deployed across wide area networks using the (public) Internet infrastructure to achieve high availability through replication. 
Networks, however, can be impacted by different adverse network conditions, ranging from accidental (e.g., a network congestion can affect the communication to and from the current leader slowing down all nodes) to intentional (e.g., a carefully crafted DDoS attack can interfere with a consensus replica group).

DDoS is a relatively simple but powerful technique to attack Internet resources~\cite{douligeris2004ddos}, preventing or limiting access to a targeted resource.
In the context of consensus, an attacker can perform a DDoS attack by carefully analyzing the traffic using traffic analysis, and attacking the leader node to degrade the performance of the system by forcing the replicas to follow the slow execution paths such as view change~\cite{nikolaou2015turtle}.

In this paper, we make use of the DDoS attack description of Spiegelman \textit{et al.}~\cite{spiegelman2019ace} to represent DDoS attacks relevant to consensus. 
We will refer to an attack that affects a consensus protocol as a \textit{delayed view change attack}. A delayed view change attack aims to degrade the performance of a consensus algorithm \textit{while} maximizing the time it takes to elect a new leader by (1) saturating the resources of leader replica and (2) avoiding a view change for the maximum possible amount of time. 
Saturating the leader replica in a consensus system slows down the entire replica set. However, leader-based consensus algorithms are configured to trigger a view change to elect a new leader when the current one becomes unresponsive for a predefined time period. If the attacker targets the leader in a way that immediately triggers a view change, then the new leader will keep the system available, foiling the attack. Hence, the attacker has to consider the trade-off between the performance loss due to the attack and the frequency at which a new leader is elected. Delayed view change attack differs from regular leader failures such that in the regular leader failures the leader node is permanently made unavailable where as in the delayed view change attack the leader node is slowed down temporarily for a time duration that is less than view change time. While the effect of permanent leader failure is widely explored in the previous work \cite{barcelona2008mencius}, we found that the effect of delayed view change attack has not been explored in the previous work.   

\subsection{Random Exponential Backoff (REB)}

REB is a mechanism that enables a set of nodes to consume a shared resource without relying on a centralized point of entry. REB emerged as a standard technique to access shared resources in Ethernet~\cite{kurose1986computer} and the DOCSIS cable network~\cite{specification2013data}. In Ethernet, when there are concurrent data transmissions in the shared data link medium, the nodes detect the collision and re-transmit the frame. To avoid further collisions, each node backs off a random amount of time, exponentially increasing the random timeout duration. 
REB became widely used when shared-access physical links were common due to its robustness, efficiency and simplicity. 

In contrast with its firm establishment in networking, REB has not been well studied in the context of consensus. Exponential timeouts have been used in consensus protocols but mainly as a method to \textit{adjust} the leader timeouts. Multi-Paxos~\cite{lamport2001paxos,van2015paxos} and Raft employ random exponential timeouts for two reasons: (1) to increase the view change timeout upon each view change and (2) to avoid two replicas concurrently issuing a new view change request. However, none of the previous work have explored REB as a leader \textit{replacement} method, and to the best of our knowledge, our work is the first attempt to leverage and thoroughly evaluate REB as a primary method of contention 
handling in consensus.
\section{Design}\label{sec:design}

In this section we first describe \sys's system model followed by the algorithm itself. Afterwards we describe how REB integrates with \sys, provide a consensus proof sketch, and finally discuss some optimizations.

\subsection{System Model}

Let $n$ denote the number of replicas and let $f$ denote the maximum number of failed nodes. We assume $n = 2f+1$ and benign crash stop failures. For simplicity we further assume that crashes are permanent although node recovery can be easily integrated into \sys using standard recovery approaches like sync-on-disk for each operation~\cite{moraru2013there,cachin2011introduction}.

We assume perfect point-to-point links between each pair of nodes, i.e., messages sent to non-failed nodes are eventually delivered~\cite{cachin2011introduction}. This is a stronger assumption than the one made in Paxos~\cite{lamport2001paxos} where messages can be dropped. However, implementing perfect point-to-point links on top of a fair-loss link abstraction is possible by using a stubborn transmission technique~\cite{cachin2011introduction}.
In practice, TCP provides reliable communication channels. We also assume that nodes are connected in a logical complete graph. 

We assume a partially-synchronous network as defined in Dwork \textit{et al.}~\cite{dwork1988consensus}. Let $R$ be an execution of the consensus algorithm, $\Delta$ be the upper bound on message transmission delay and GST be the global stabilization time. The partial synchrony assumption states that for every $R$ there is an unknown time GST such that $\Delta$ holds in [GST, $\infty$); once GST is reached, each message sent by process $p_i$ is delivered by process $p_j$ within a known maximum time bound of $\Delta$. This assumption is necessary to guarantee the liveness of Baxos in the light of the FLP impossibility result~\cite{fischer1985impossibility}.

\subsection{The \sys Algorithm}

At its core, \sys uses the same logic as the Synod core of the Paxos protocol ({\it \synod}) \cite{lamport1989part}, where each replica can propose values. However, \synod fails to achieve liveness if there are concurrent proposals for the same consensus instance. \sys addresses this liveness issue by using REB: if there are concurrent requests for the same consensus instance, \sys replicas {\it back off} for a random amount of time to prevent further collisions. This ensures that one proposer succeeds in committing their value for the consensus instance within a few retries.

\begin{figure*}
    
	\centering
	\resizebox{\textwidth}{!}{

\begin{tikzpicture}[x=0.75pt,y=0.75pt,yscale=-1,xscale=1]

\draw    (75.52,95.06) -- (919.5,87.32) ;
\draw    (76.55,168.02) -- (919.5,162.5) ;
\draw    (76.55,240.99) -- (919.5,235.46) ;
\draw    (294.35,67.42) -- (294.35,269.73) ;
\draw    (80.66,97.27) -- (115.12,240.15) ;
\draw [shift={(115.59,242.09)}, rotate = 256.44] [color={rgb, 255:red, 0; green, 0; blue, 0 }  ][line width=0.75]    (10.93,-3.29) .. controls (6.95,-1.4) and (3.31,-0.3) .. (0,0) .. controls (3.31,0.3) and (6.95,1.4) .. (10.93,3.29)   ;
\draw    (80.66,97.27) -- (111.69,166.2) ;
\draw [shift={(112.51,168.02)}, rotate = 245.76999999999998] [color={rgb, 255:red, 0; green, 0; blue, 0 }  ][line width=0.75]    (10.93,-3.29) .. controls (6.95,-1.4) and (3.31,-0.3) .. (0,0) .. controls (3.31,0.3) and (6.95,1.4) .. (10.93,3.29)   ;
\draw    (112.51,168.02) -- (151.61,96.81) ;
\draw [shift={(152.58,95.06)}, rotate = 478.77] [color={rgb, 255:red, 0; green, 0; blue, 0 }  ][line width=0.75]    (10.93,-3.29) .. controls (6.95,-1.4) and (3.31,-0.3) .. (0,0) .. controls (3.31,0.3) and (6.95,1.4) .. (10.93,3.29)   ;
\draw    (115.59,242.09) -- (152.09,97) ;
\draw [shift={(152.58,95.06)}, rotate = 464.12] [color={rgb, 255:red, 0; green, 0; blue, 0 }  ][line width=0.75]    (10.93,-3.29) .. controls (6.95,-1.4) and (3.31,-0.3) .. (0,0) .. controls (3.31,0.3) and (6.95,1.4) .. (10.93,3.29)   ;
\draw    (179.29,95.06) -- (213.75,237.94) ;
\draw [shift={(214.22,239.88)}, rotate = 256.44] [color={rgb, 255:red, 0; green, 0; blue, 0 }  ][line width=0.75]    (10.93,-3.29) .. controls (6.95,-1.4) and (3.31,-0.3) .. (0,0) .. controls (3.31,0.3) and (6.95,1.4) .. (10.93,3.29)   ;
\draw    (179.29,95.06) -- (210.32,163.99) ;
\draw [shift={(211.14,165.81)}, rotate = 245.76999999999998] [color={rgb, 255:red, 0; green, 0; blue, 0 }  ][line width=0.75]    (10.93,-3.29) .. controls (6.95,-1.4) and (3.31,-0.3) .. (0,0) .. controls (3.31,0.3) and (6.95,1.4) .. (10.93,3.29)   ;
\draw    (211.14,165.81) -- (250.24,94.6) ;
\draw [shift={(251.2,92.85)}, rotate = 478.77] [color={rgb, 255:red, 0; green, 0; blue, 0 }  ][line width=0.75]    (10.93,-3.29) .. controls (6.95,-1.4) and (3.31,-0.3) .. (0,0) .. controls (3.31,0.3) and (6.95,1.4) .. (10.93,3.29)   ;
\draw    (214.22,239.88) -- (250.72,94.79) ;
\draw [shift={(251.2,92.85)}, rotate = 464.12] [color={rgb, 255:red, 0; green, 0; blue, 0 }  ][line width=0.75]    (10.93,-3.29) .. controls (6.95,-1.4) and (3.31,-0.3) .. (0,0) .. controls (3.31,0.3) and (6.95,1.4) .. (10.93,3.29)   ;
\draw    (251.2,92.85) -- (285.66,235.73) ;
\draw [shift={(286.13,237.67)}, rotate = 256.44] [color={rgb, 255:red, 0; green, 0; blue, 0 }  ][line width=0.75]    (10.93,-3.29) .. controls (6.95,-1.4) and (3.31,-0.3) .. (0,0) .. controls (3.31,0.3) and (6.95,1.4) .. (10.93,3.29)   ;
\draw    (251.2,92.85) -- (282.78,167.29) ;
\draw [shift={(283.56,169.13)}, rotate = 247.01] [color={rgb, 255:red, 0; green, 0; blue, 0 }  ][line width=0.75]    (10.93,-3.29) .. controls (6.95,-1.4) and (3.31,-0.3) .. (0,0) .. controls (3.31,0.3) and (6.95,1.4) .. (10.93,3.29)   ;
\draw    (306.68,95.06) -- (341.14,237.94) ;
\draw [shift={(341.61,239.88)}, rotate = 256.44] [color={rgb, 255:red, 0; green, 0; blue, 0 }  ][line width=0.75]    (10.93,-3.29) .. controls (6.95,-1.4) and (3.31,-0.3) .. (0,0) .. controls (3.31,0.3) and (6.95,1.4) .. (10.93,3.29)   ;
\draw    (306.68,95.06) -- (337.71,163.99) ;
\draw [shift={(338.53,165.81)}, rotate = 245.76999999999998] [color={rgb, 255:red, 0; green, 0; blue, 0 }  ][line width=0.75]    (10.93,-3.29) .. controls (6.95,-1.4) and (3.31,-0.3) .. (0,0) .. controls (3.31,0.3) and (6.95,1.4) .. (10.93,3.29)   ;
\draw    (370.38,168.02) -- (401.4,236.95) ;
\draw [shift={(402.22,238.78)}, rotate = 245.76999999999998] [color={rgb, 255:red, 0; green, 0; blue, 0 }  ][line width=0.75]    (10.93,-3.29) .. controls (6.95,-1.4) and (3.31,-0.3) .. (0,0) .. controls (3.31,0.3) and (6.95,1.4) .. (10.93,3.29)   ;
\draw    (370.38,168.02) -- (405.48,93.55) ;
\draw [shift={(406.33,91.74)}, rotate = 475.24] [color={rgb, 255:red, 0; green, 0; blue, 0 }  ][line width=0.75]    (10.93,-3.29) .. controls (6.95,-1.4) and (3.31,-0.3) .. (0,0) .. controls (3.31,0.3) and (6.95,1.4) .. (10.93,3.29)   ;
\draw    (555.3,63) -- (555.3,265.31) ;
\draw    (427.91,93.95) -- (462.37,236.83) ;
\draw [shift={(462.84,238.78)}, rotate = 256.44] [color={rgb, 255:red, 0; green, 0; blue, 0 }  ][line width=0.75]    (10.93,-3.29) .. controls (6.95,-1.4) and (3.31,-0.3) .. (0,0) .. controls (3.31,0.3) and (6.95,1.4) .. (10.93,3.29)   ;
\draw    (427.91,93.95) -- (459.47,166.19) ;
\draw [shift={(460.27,168.02)}, rotate = 246.4] [color={rgb, 255:red, 0; green, 0; blue, 0 }  ][line width=0.75]    (10.93,-3.29) .. controls (6.95,-1.4) and (3.31,-0.3) .. (0,0) .. controls (3.31,0.3) and (6.95,1.4) .. (10.93,3.29)   ;
\draw    (491.6,168.02) -- (526.71,93.55) ;
\draw [shift={(527.56,91.74)}, rotate = 475.24] [color={rgb, 255:red, 0; green, 0; blue, 0 }  ][line width=0.75]    (10.93,-3.29) .. controls (6.95,-1.4) and (3.31,-0.3) .. (0,0) .. controls (3.31,0.3) and (6.95,1.4) .. (10.93,3.29)   ;
\draw    (491.6,168.02) -- (522.63,236.95) ;
\draw [shift={(523.45,238.78)}, rotate = 245.76999999999998] [color={rgb, 255:red, 0; green, 0; blue, 0 }  ][line width=0.75]    (10.93,-3.29) .. controls (6.95,-1.4) and (3.31,-0.3) .. (0,0) .. controls (3.31,0.3) and (6.95,1.4) .. (10.93,3.29)   ;
\draw    (582.01,93.95) -- (616.47,236.83) ;
\draw [shift={(616.94,238.78)}, rotate = 256.44] [color={rgb, 255:red, 0; green, 0; blue, 0 }  ][line width=0.75]    (10.93,-3.29) .. controls (6.95,-1.4) and (3.31,-0.3) .. (0,0) .. controls (3.31,0.3) and (6.95,1.4) .. (10.93,3.29)   ;
\draw    (582.01,93.95) -- (613.04,162.88) ;
\draw [shift={(613.86,164.71)}, rotate = 245.76999999999998] [color={rgb, 255:red, 0; green, 0; blue, 0 }  ][line width=0.75]    (10.93,-3.29) .. controls (6.95,-1.4) and (3.31,-0.3) .. (0,0) .. controls (3.31,0.3) and (6.95,1.4) .. (10.93,3.29)   ;
\draw    (651.87,165.81) -- (686.98,91.34) ;
\draw [shift={(687.83,89.53)}, rotate = 475.24] [color={rgb, 255:red, 0; green, 0; blue, 0 }  ][line width=0.75]    (10.93,-3.29) .. controls (6.95,-1.4) and (3.31,-0.3) .. (0,0) .. controls (3.31,0.3) and (6.95,1.4) .. (10.93,3.29)   ;
\draw    (651.87,165.81) -- (682.9,234.74) ;
\draw [shift={(683.72,236.56)}, rotate = 245.76999999999998] [color={rgb, 255:red, 0; green, 0; blue, 0 }  ][line width=0.75]    (10.93,-3.29) .. controls (6.95,-1.4) and (3.31,-0.3) .. (0,0) .. controls (3.31,0.3) and (6.95,1.4) .. (10.93,3.29)   ;
\draw    (737.14,89.53) -- (771.61,232.41) ;
\draw [shift={(772.07,234.35)}, rotate = 256.44] [color={rgb, 255:red, 0; green, 0; blue, 0 }  ][line width=0.75]    (10.93,-3.29) .. controls (6.95,-1.4) and (3.31,-0.3) .. (0,0) .. controls (3.31,0.3) and (6.95,1.4) .. (10.93,3.29)   ;
\draw    (737.14,89.53) -- (771.73,162.9) ;
\draw [shift={(772.59,164.71)}, rotate = 244.76] [color={rgb, 255:red, 0; green, 0; blue, 0 }  ][line width=0.75]    (10.93,-3.29) .. controls (6.95,-1.4) and (3.31,-0.3) .. (0,0) .. controls (3.31,0.3) and (6.95,1.4) .. (10.93,3.29)   ;
\draw    (772.59,164.71) -- (808.21,89.13) ;
\draw [shift={(809.06,87.32)}, rotate = 475.23] [color={rgb, 255:red, 0; green, 0; blue, 0 }  ][line width=0.75]    (10.93,-3.29) .. controls (6.95,-1.4) and (3.31,-0.3) .. (0,0) .. controls (3.31,0.3) and (6.95,1.4) .. (10.93,3.29)   ;
\draw    (772.07,234.35) -- (808.57,89.26) ;
\draw [shift={(809.06,87.32)}, rotate = 464.12] [color={rgb, 255:red, 0; green, 0; blue, 0 }  ][line width=0.75]    (10.93,-3.29) .. controls (6.95,-1.4) and (3.31,-0.3) .. (0,0) .. controls (3.31,0.3) and (6.95,1.4) .. (10.93,3.29)   ;
\draw    (823.44,89.53) -- (857.9,232.41) ;
\draw [shift={(858.37,234.35)}, rotate = 256.44] [color={rgb, 255:red, 0; green, 0; blue, 0 }  ][line width=0.75]    (10.93,-3.29) .. controls (6.95,-1.4) and (3.31,-0.3) .. (0,0) .. controls (3.31,0.3) and (6.95,1.4) .. (10.93,3.29)   ;
\draw    (823.44,89.53) -- (854.47,158.46) ;
\draw [shift={(855.29,160.28)}, rotate = 245.76999999999998] [color={rgb, 255:red, 0; green, 0; blue, 0 }  ][line width=0.75]    (10.93,-3.29) .. controls (6.95,-1.4) and (3.31,-0.3) .. (0,0) .. controls (3.31,0.3) and (6.95,1.4) .. (10.93,3.29)   ;
\draw    (855.29,160.28) -- (875.41,123.65) -- (894.39,89.07) ;
\draw [shift={(895.36,87.32)}, rotate = 478.77] [color={rgb, 255:red, 0; green, 0; blue, 0 }  ][line width=0.75]    (10.93,-3.29) .. controls (6.95,-1.4) and (3.31,-0.3) .. (0,0) .. controls (3.31,0.3) and (6.95,1.4) .. (10.93,3.29)   ;
\draw    (858.37,234.35) -- (894.87,89.26) ;
\draw [shift={(895.36,87.32)}, rotate = 464.12] [color={rgb, 255:red, 0; green, 0; blue, 0 }  ][line width=0.75]    (10.93,-3.29) .. controls (6.95,-1.4) and (3.31,-0.3) .. (0,0) .. controls (3.31,0.3) and (6.95,1.4) .. (10.93,3.29)   ;

\draw (3.47,89.32) node [anchor=north west][inner sep=0.75pt]   [align=left] {Replica 1};
\draw (2.44,156.76) node [anchor=north west][inner sep=0.75pt]   [align=left] {Replica 2};
\draw (1.42,230.83) node [anchor=north west][inner sep=0.75pt]   [align=left] {Replica 3};
\draw (79.58,255.57) node [anchor=north west][inner sep=0.75pt]   [align=left] {\begin{minipage}[lt]{140.51pt}\setlength\topsep{0pt}
\begin{center}
 \ \ (a) Synod Paxos and Baxos \\without contention
\end{center}

\end{minipage}};
\draw (317.02,255.78) node [anchor=north west][inner sep=0.75pt]   [align=left] {\begin{minipage}[lt]{149.58pt}\setlength\topsep{0pt}
\begin{center}
 \ \ (b) Contention in Synod Paxos\\can lead to livelock.
\end{center}

\end{minipage}};
\draw (580.09,253.68) node [anchor=north west][inner sep=0.75pt]  [font=\fontsize{1.07em}{1.28em}\selectfont] [align=left] { \ \ (c) Baxos handles contention using backoff};
\draw (825.03,198.4) node [anchor=north west][inner sep=0.75pt]  [font=\small] [align=left] {Propose};
\draw (264.96,197.29) node [anchor=north west][inner sep=0.75pt]  [font=\small] [align=left] {Learn};
\draw (761.33,126.54) node [anchor=north west][inner sep=0.75pt]  [font=\small] [align=left] {Promise};
\draw (773.66,212.77) node [anchor=north west][inner sep=0.75pt]  [font=\small] [align=left] {Promise};
\draw (822.98,113.28) node [anchor=north west][inner sep=0.75pt]  [font=\small] [align=left] {Propose};
\draw (175.74,190.66) node [anchor=north west][inner sep=0.75pt]  [font=\small] [align=left] {Propose};
\draw (174.71,115.49) node [anchor=north west][inner sep=0.75pt]  [font=\small] [align=left] {Propose};
\draw (203.38,130.96) node [anchor=north west][inner sep=0.75pt]  [font=\small] [align=left] {Accept};
\draw (218.79,181.82) node [anchor=north west][inner sep=0.75pt]  [font=\small] [align=left] {Accept};
\draw (872.19,124.49) node [anchor=north west][inner sep=0.75pt]  [font=\small] [align=left] {Accept};
\draw (871.17,180.71) node [anchor=north west][inner sep=0.75pt]  [font=\small] [align=left] {Accept};
\draw (258.79,126.54) node [anchor=north west][inner sep=0.75pt]  [font=\small] [align=left] {Learn};
\draw (674.12,145.92) node [anchor=north west][inner sep=0.75pt]  [font=\large,color={rgb, 255:red, 208; green, 2; blue, 27 }  ,opacity=1 ] [align=left] {\textbf{{\scriptsize Replica 2}}\\\textbf{{\scriptsize Backs off}}};
\draw (792.28,145.81) node [anchor=north west][inner sep=0.75pt]  [font=\large,color={rgb, 255:red, 65; green, 117; blue, 5 }  ,opacity=1 ] [align=left] {\textbf{{\scriptsize Replica 1}}\\\textbf{{\scriptsize Proposes}}};
\draw (390.5,169.66) node [anchor=north west][inner sep=0.75pt]  [font=\footnotesize,color={rgb, 255:red, 208; green, 2; blue, 27 }  ,opacity=1 ] [align=left] {\textbf{Livelock}};
\draw (116.66,186.77) node [anchor=north west][inner sep=0.75pt]  [font=\small] [align=left] {Promise};
\draw (129.66,142.77) node [anchor=north west][inner sep=0.75pt]  [font=\small] [align=left] {Promise};
\draw (319.66,185.77) node [anchor=north west][inner sep=0.75pt]  [font=\small] [align=left] {Prepare};
\draw (362.66,133.77) node [anchor=north west][inner sep=0.75pt]  [font=\small] [align=left] {Prepare};
\draw (384.66,202.77) node [anchor=north west][inner sep=0.75pt]  [font=\small] [align=left] {Prepare};
\draw (426.66,108.77) node [anchor=north west][inner sep=0.75pt]  [font=\small] [align=left] {Prepare};
\draw (312.66,114.77) node [anchor=north west][inner sep=0.75pt]  [font=\small] [align=left] {Prepare};
\draw (69.66,201.77) node [anchor=north west][inner sep=0.75pt]  [font=\small] [align=left] {Prepare};
\draw (449.66,198.77) node [anchor=north west][inner sep=0.75pt]  [font=\small] [align=left] {Prepare};
\draw (58.66,128.77) node [anchor=north west][inner sep=0.75pt]  [font=\small] [align=left] {Prepare};
\draw (484.66,125.77) node [anchor=north west][inner sep=0.75pt]  [font=\small] [align=left] {Prepare};
\draw (493.6,171.02) node [anchor=north west][inner sep=0.75pt]  [font=\small] [align=left] {Prepare};
\draw (582.66,195.77) node [anchor=north west][inner sep=0.75pt]  [font=\small] [align=left] {Prepare};
\draw (590.66,122.77) node [anchor=north west][inner sep=0.75pt]  [font=\small] [align=left] {Prepare};
\draw (649.66,195.77) node [anchor=north west][inner sep=0.75pt]  [font=\small] [align=left] {Prepare};
\draw (656.66,121.77) node [anchor=north west][inner sep=0.75pt]  [font=\small] [align=left] {Prepare};
\draw (732.66,190.77) node [anchor=north west][inner sep=0.75pt]  [font=\small] [align=left] {Prepare};
\draw (716.66,110.77) node [anchor=north west][inner sep=0.75pt]  [font=\small] [align=left] {Prepare};

\end{tikzpicture}

    }
	\caption{A space time diagram showing the message flow of Synod Paxos and Baxos with and without contention}
	\label{fig:Baxos_algo}

\end{figure*}
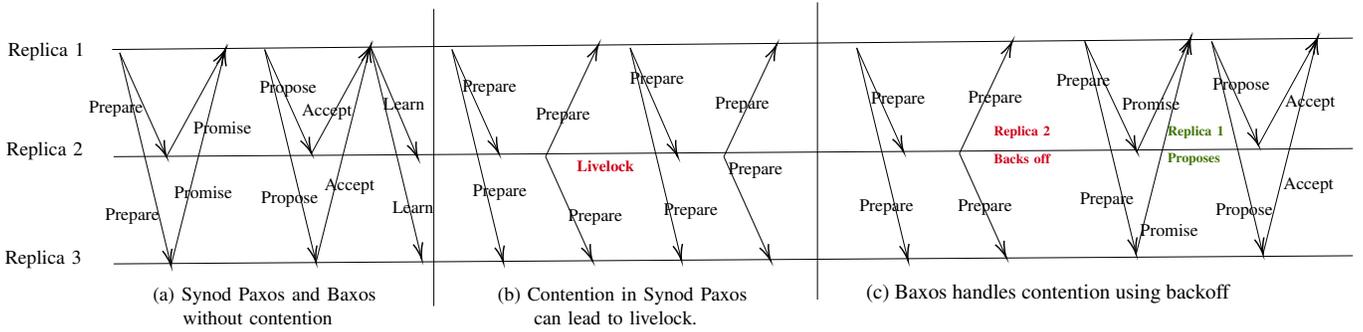

We use the term {\it try} to denote the concept of {\it Ballot} number in \synod, and the term {\it choice} to indicate a consensus instance. A sequence of {\it choice} elements make the replicated log. 
As in Paxos, each replica can take on the role as an {\it Acceptor}, {\it Proposer}, and {\it Learner}~\cite{lamport2001paxos}.

\synod and single-choice \sys consist of the following two phases, see Figure~\ref{fig:Baxos_algo}\,(a):

\paragraph{Prepare-Promise.} A node which receives a new command from the upper layer takes on the role of Proposer and initiates consensus by broadcasting a Prepare message to all Acceptors. The Prepare message contains a {\it proposed\_try} number, which keeps track of the current {\it try} number. Acceptors send a Promise message to the Proposer, if they have not accepted any Prepare message with a higher or equal {\it try} number than the {\it proposed\_try} received in the Prepare message. To inform the Proposer about any previously accepted value, Acceptors piggyback the highest {\it try} for which they last accepted a value, and the corresponding value. If the Proposer manages to collect Promise messages from a majority, i.e., $f+1$ or more, of Acceptors, then it selects the previously accepted value corresponding to the highest received previously accepted {\it try} number, chosen from the received set of Promise messages. If all Promise messages indicate that there is no such previously accepted value, then the proposer  selects the received command $c$ from the upper layer as the value to propose. Let {\it proposed\_value} denote this selected value.

\paragraph{Propose-Accept.} Upon successfully collecting Promise messages from a majority of Acceptors, the Proposer broadcasts a Propose message piggybacked with the {\it proposed\_try} and the {\it proposed\_value}. An Acceptor accepts a Propose message, if the {\it try} number in the Propose message is greater than or equal to the highest {\it try} number that it promised. Upon accepting a Propose message from the Proposer,  Acceptors update their {\it accepted\_try} and {\it accepted\_value} variables with {\it proposed\_try} and the {\it proposed\_value}, respectively. and send an Accept message to the Proposer. The Proposer, upon receiving Accept messages from a majority of Acceptors, decides on that value and informs the upper layer about the decision. Finally, the proposer broadcasts a Learn message to inform Learners about the decision.

\subsubsection{The Liveness Challenge}

The above two-phase algorithm is the core of \synod, and it achieves obstruction-free but not lock-free termination: If there are multiple concurrent Proposers, then the above algorithm fails to terminate. An example execution where the termination property is not achieved is depicted in Figure~\ref{fig:Baxos_algo}\,(b) , where replica 1 and replica 2 concurrently send the Prepare messages, without making any progress. In \synod, upon learning contention, the Proposer retries the Prepare-Promise phase with a {\it proposed\_try} that is strictly greater than its previous {\it proposed\_try} and {\it promised\_try}. However, immediately retrying phase 1 causes further contention.

Addressing contention is where \sys differs from \synod: whereas \synod does not implement a mechanism to deal with contention, \sys uses REB to address contention. To avoid contention and achieve lock-free termination, the Proposer in \sys backs off for a random exponential timeout before retrying again. Figure~\ref{fig:Baxos_algo}\,(c) illustrates how \sys backs off to handle contention.

REB is appealing as a method of handling contention in \synod due to three main guarantees of REB: (1) robustness, (2) high throughput and (3) resource utilization efficiency \cite{bender2016scale} as studied in the networking literature. REB enables appointing nodes in non-conflicting timeouts, so that there is only one node utilizing the shared recourse at a given time interval. In this paper, we ask the question "can REB bring the same advantages to the domain of consensus protocols?". In the next section we present our random back off scheme and explain why it achieves lock-free termination.

\subsection{REB in Baxos}

We aim to achieve two objectives from our REB scheme: (1) Provide lock-free termination by concluding a single Proposer for a consensus choice with asymptotically logarithmic number of failed proposals (retries) and (2) adapt to changing wide area network conditions such as variable latency.

We propose a REB scheme, called Baxos REB, that achieves the two goals above. We did not employ the existing REB schemes proposed in the networking literature, because they were designed under different assumptions, such as synchronous frame-transmission times, that do not apply in consensus, where the inter-replica latency varies significantly.

In the Baxos REB scheme, upon facing $l$ retries, each node first selects a number $k \in (0,1) \subseteq \mathbb{Q}$ uniformly at random. Then each node backs off for $k \times 2^{l} \times 2\times RTT$ time period where RTT is the maximum network round trip time between any pair of replicas (network diameter) (note that $\Delta$ is the upper bound of RTT/2). Note that we use $2 \times RTT$ in our backoff time calculation, because there are two network round trips to commit a single request (Prepare-Promise and Propose-Accept) and to allow another proposer to successfully propose a command, other replicas should backoff a minimum of $2 \times RTT$.  Upon successfully proposing a value, $l$ is  decreased by one. As shown in Section~\ref{correctness}, Baxos REB ensures that eventually (after reaching GST) there exists only one Proposer for a sufficiently large time period $4 \times \Delta$ (note that when GST is reached RTT = 2 $\times \Delta$), such that a decision is made.

Baxos REB scheme achieves the two objectives we intended (liveness and robustness). Baxos REB achieves liveness due to our use of a continuous interval to choose a random value from. Using a continuous interval enables Baxos to have low probability of two nodes selecting the same value (compared to selecting a value from a discrete set in binary REB in networking \cite{kurose1986computer}). Baxos REB scheme achieves robustness to changing wide area network conditions by dynamically monitoring the RTT for the back off time calculation. This enables Baxos to adapt its backoff time calculations with respect to changing wide area network conditions. Had we used the binary REB~\cite{kurose1986computer}, which is widely used in networking, due to its use of constant frame transmission time to calculate the timeout, we would not be able to adapt to changing network dynamics.

\subsection{Consensus Proof Sketch}\label{correctness}

We now provide a proof sketch for single-choice Baxos, which satisfies the following four consensus properties~\cite{cachin2011introduction}:

\begin{enumerate}[noitemsep,topsep=0pt,parsep=0pt,partopsep=0pt]
\item {\it Validity}: If any node decides on a value $v$, then $v$ should have been previously proposed by some proposer.
\item {\it Termination}: If a correct process proposes a value, then that process eventually decides.
\item {\it Agreement}: No two nodes decide differently.
\item {\it Integrity}: No node decides more than once.
\end{enumerate}

Validity, agreement, and integrity directly follow from the Synod Paxos proofs because we use the same core as Synod Paxos. 
Termination is derived using our REB scheme. This section focuses on the termination proof sketch for single-choice Baxos.

Termination of Baxos holds only after the GST is reached, and when there is an upper bound $\Delta$ on the message transmission time between any pair of nodes. If there is only a single proposer for a run of Baxos, the protocol trivially terminates, hence we focus on the case with multiple contending proposers.

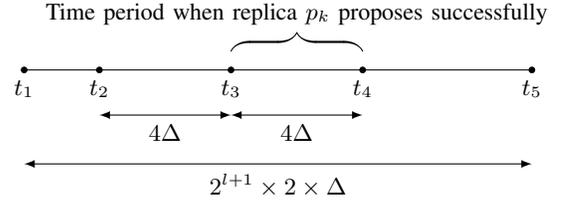
\begin{figure}[ht]
	\centering
    \begin{tikzpicture}[>=latex, font=\small]
        \tikzset{
        DOT/.style={draw,circle,fill=black,inner sep=0.75pt}}
        \node (zero) at (0.00, 0.00) {};
        \node (t1) at ($(zero) + ( 0.00, 0.00)$) {$t_1$};
        \node (t2) at ($(t1) + ( 1.00, 0.00)$) {$t_2$};
        \node (t3) at ($(t2) + ( 1.75, 0.00)$) {$t_3$};
        \node (t4) at ($(t3) + ( 1.75, 0.00)$) {$t_4$};
        \node (t5) at ($(t4) + ( 2.25, 0.00)$) {$t_5$};
        \node[DOT] (d1) at ($(t1) + ( 0.00, 0.25)$) {};
        \node[DOT] (d2) at ($(t2) + ( 0.00, 0.25)$) {};
        \node[DOT] (d3) at ($(t3) + ( 0.00, 0.25)$) {};
        \node[DOT] (d4) at ($(t4) + ( 0.00, 0.25)$) {};
        \node[DOT] (d5) at ($(t5) + ( 0.00, 0.25)$) {};
        \draw[-] ($(t1) + ( 0.0, 0.25)$) -- ($(t5) + ( 0.0, 0.25)$);
        \draw[<->] ($(t2) + ( 0.0,-0.35)$) -- ($(t3) + ( 0.0,-0.35)$) node [pos=0.5,below] {$4\Delta$};
        \draw[<->] ($(t3) + ( 0.0,-0.35)$) -- ($(t4) + ( 0.0,-0.35)$) node [pos=0.5,below] {$4\Delta$};
        \draw[<->] ($(t1) + ( 0.0,-1.00)$) -- ($(t5) + ( 0.0,-1.00)$) node [pos=0.5,below] {$2^{l+1} \times 2 \times \Delta$};
        \draw [decorate, thick, decoration = {calligraphic brace, amplitude=7pt}] ($(t3) + ( 0.0, 0.5)$) -- ($(t4) + ( 0.0, 0.5)$);
        \node at ($(t3) + ( 0.875, 1.0)$) {Time period when replica $p_k$ proposes successfully};
    \end{tikzpicture}
	\caption{Illustration of Baxos termination.}
	\label{fig:contbackoff}
\end{figure}

If there are multiple competing proposals from different proposers, each node backs off over a time period of length $k \times2^{l+1} \times 2 \times \Delta$, since $RTT = 2\times \Delta$ after GST is reached, where $k \in (0,1) \subseteq \mathbb{Q}$ and $l$ is the number of retries. For termination to hold, we need to show that with high probability there exists a time interval of length $2 \times 2 \times \Delta$ in which only a single replica stops backing off and makes its proposal. Figure~\ref{fig:contbackoff} illustrates this scenario.

Assume that there are $p$ replicas which compete to propose a value and that each has done already $l$ retries. Assume that all replicas start backing off at time $t_1$.
Let $t_5$ denote the time at which the last node finishes backing off, then the time interval $(t_1:t_5)$ has a maximum length of $2^{l+2}\times \Delta$. Depending on $k$, replica $p_k$ can stop backing off at any time in $(t_1:t_5)$ and start its proposal phase.
Let $t_3$ denote the time at which proposer $p_k$ stops backing off and starts proposing and let $t_4$ denote the time at which $p_k$ successfully decides. The interval $(t_3:t_4)$ is of length $4\Delta$ which is the duration a proposal requires to complete successfully. Let $(t_2:t_3)$ denote the interval of length $4\Delta$ before $(t_3:t_4)$.
For $p_k$ to terminate, no other replica should propose in $(t_2:t_4)$ of length $2 \times 4 \times \Delta$ since any replica which stops its back-off and proposes after $t_2$ will make its proposal in $(t_3:t_4)$ putting it into conflict with $p_k$'s proposal.
Thus the probability that replica $p_k$ is the only proposer in $(t_{2}:t_{4})$ is equal to the probability that all other $p-1$ proposers finish their back-offs and proposals in the intervals ($t_{1}:t_{2}$) and ($t_{4}:t_{5}$). This probability is given in Equation~\ref{continuous}. 

\begin{equation} \label{continuous}
\left(\frac{2^{l+2}\Delta-8\Delta}{2^{l+2}\Delta}\right)^{p-1} = \left(1- \frac{1}{2^{l-1}}\right)^{p-1}
\end{equation}

If the value of $l$ is large enough, this probability approaches $1$. Hence node $p_{k}$ eventually succeeds in proposing its value and thus decides.

This proof sketch for termination assumed that each contending proposer starts to backoff at the same time $t_{1}$ and that each contending proposer has experienced the same number of retries $l$. In our experiments, we observed that different proposers start to backoff at different times. For the simplicity of our proof we can let the adversary manipulate the delivery times of messages such that each node starts the backoff timer from the beginning of the synchronized period even if the conflict of replicas is detected at some point $t_{1}+t$ where $t<\Delta$ (otherwise a new period starts). Additionally,  different replicas have different $l$, but since the backoffs are exponentially increasing it is obvious that eventually all replicas will reach the same $l$ (this is the same reasoning as the view-change timeouts backoff in Raft).

\subsection{One-Round Trip Optimization}
In the absence of leader failures and network partitions, Multi-Paxos consumes a single network round trip time to commit a single client request (when the client to leader network round trip time is not considered). This is possible in Multi-Paxos because the leader node runs the Prepare-Promise phase for a sequence of consensus instances, and thereafter, only the leader proposes the commands.

In contrast, Baxos consumes two network round trip delays to commit a single client request, which is a significant drawback. To address this drawback, we apply a classic message piggybacking technique, where the Prepare message for the choice $i$ is piggybacked in the Propose message of choice $i-1$ similar to \cite{kogias2016enhancing}. Since the Prepare-Promise phase of choice $i$ does not depend on the Propose-Accept phase of choice $i-1$, the Prepare message for choice $i$ can be piggybacked on the choice $i-1$ Propose message. 
This optimization enables Baxos to commit a request in a single network round trip time, when successive client requests are proposed by the same Proposer. When multiple Proposers propose concurrently (contention), this optimization does not deliver any performance benefit. Given the nature of user interacting web services where a client sends back to back requests in a partly-open system \cite{schroeder2006open}, this design decision seems a reasonable choice to achieve performance that is comparable to Raft and Multi-Paxos.
\section{Implementation}\label{sec:impl}

We implemented Baxos, Multi-Paxos, and Raft using Golang version 1.15.2. We decided to re-implement Multi-Paxos and Raft in order to have a common framework to compare the performance of these protocols. 
We cross-validated our framework implementation by comparing the Multi-Paxos results we obtained against the existing Multi-Paxos implementation\footnote{https://github.com/efficient/epaxos/} by running the experiments using the same setup and workload.

For each consensus algorithm, we used Protobuf encoding~\cite{protobuf} and gRPC~\cite{grpc} for message serialization and RPC. We implemented all the attack scenarios we present in this chapter. We did not implement snapshot and replica reconfiguration, which are outside the scope of this chapter.
\section{Evaluation}\label{sec:eval}

The goal of this evaluation is to answer following questions.

\begin{itemize}
\item How robust is \sys against delayed view change attacks in the wide-area networks?
\item  What is the performance overhead of \sys during failure-free synchronous periods in the wide-area networks?
\item How efficient is \sys in utilizing bandwidth across replicas in wide-area networks?
\item How does \sys scale with increasing replica count in the wide-area networks?
\end{itemize}

\subsection{Experimental Setup} \label{sec:setup}

We conducted our experiments using c5d.4xlarge instances (16 virtual CPUs, 32GB memory, and up to 10 Gbps network bandwidth), running Ubuntu Linux 20.04.3 LTS. 
Each AWS location has a single replica and a single client. Unless mentioned otherwise, we experiment with five consensus replicas and five client replicas ($n=5$) located in five geographically separated Amazon data centers in N. Virginia, Ireland, N. California, Tokyo, and HongKong.


In \sys, a client sends requests to the consensus replica in the same data-center as the client; if the server in the same location has failed, then the clients send requests to a randomly chosen replica in a different data-center. 
In Multi-Paxos and Raft, clients send requests to the leader replica. Clients generate requests simultaneously and measure the execution latency for each request. Each experiment was run for 1 minute and was repeated 10 times. 
We found that longer experiments do not significantly affect the performance results. To amortize the cost of the wide-area network delays, we follow the standard practice of using batching in the replica side with a maximum batch time of 5ms. This resulted in batches of size in the range (5,000, 10,000) requests.

We measure the latency on the client side starting from when a new request is sent by a client until the client receives the response. We set the client request timeout to 8 seconds and requests that took longer than 8s were treated as failed. We measure the throughput on the client side as the ratio of the number of successfully committed requests, excluding failed and timeout requests, and the time duration of the experiment. In the tests where we depict the throughput as a function of time, we aggregate the number of committed requests in one second intervals.

For the delayed view change attack performance results, we use the experimental approach of Spiegelman \textit{et al}. \cite{spiegelman2019ace}. We changed the transmission delay and packet loss of the replicas using NetEm~\cite{netem}.

\subsection{Workload} \label{sec:workload}

We use a combination of standard and synthetic micro benchmarks. Our synthetic micro benchmark consists of a configurable service time, configurable request and response sizes. We use the YCSB-A~\cite{cooper2010benchmarking} workload with the Redis~\cite{carlson2013redis} key value store as the standard benchmark.

Each command in the micro benchmark consists of $p$ bytes of payload and a unique request identifier. All client requests (reads and writes) are totally ordered in \sys. When a server receives a request, it uses consensus to totally order it, and upon committing and executing, sends a response to client with $q$ bytes with the unique request identifier. After each experiment, we use the replica logs to verify that each replica learns the same sequence of requests. We use open loop model~\cite{schroeder2006open} based on the Poisson arrival of client requests for both YCSB-A workload and the synthetic workload.

\subsection{DDoS Performance} \label{sec:eval_attack}

\begin{figure*}
    \vspace{-8mm}
    \begin{subfigure}[b]{0.5\textwidth}
      \centering
      \includegraphics[width=\textwidth]{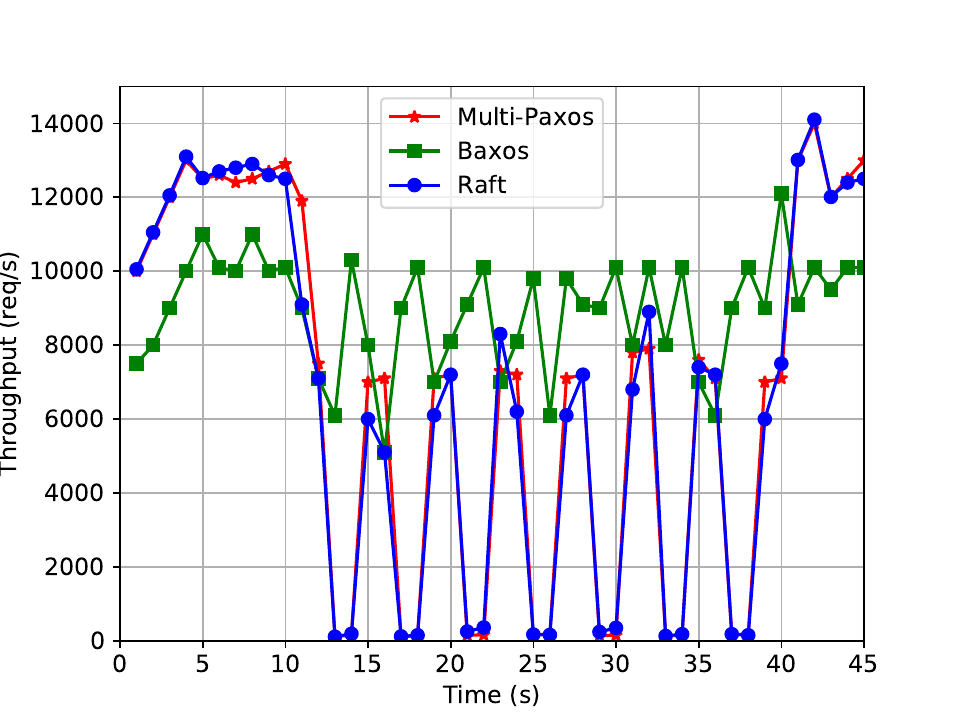}
      \caption{Throughput during delay attack}
      \vspace{-1mm}
    \end{subfigure}
    \begin{subfigure}[b]{0.5\textwidth}
      \centering
      \includegraphics[width=\textwidth]{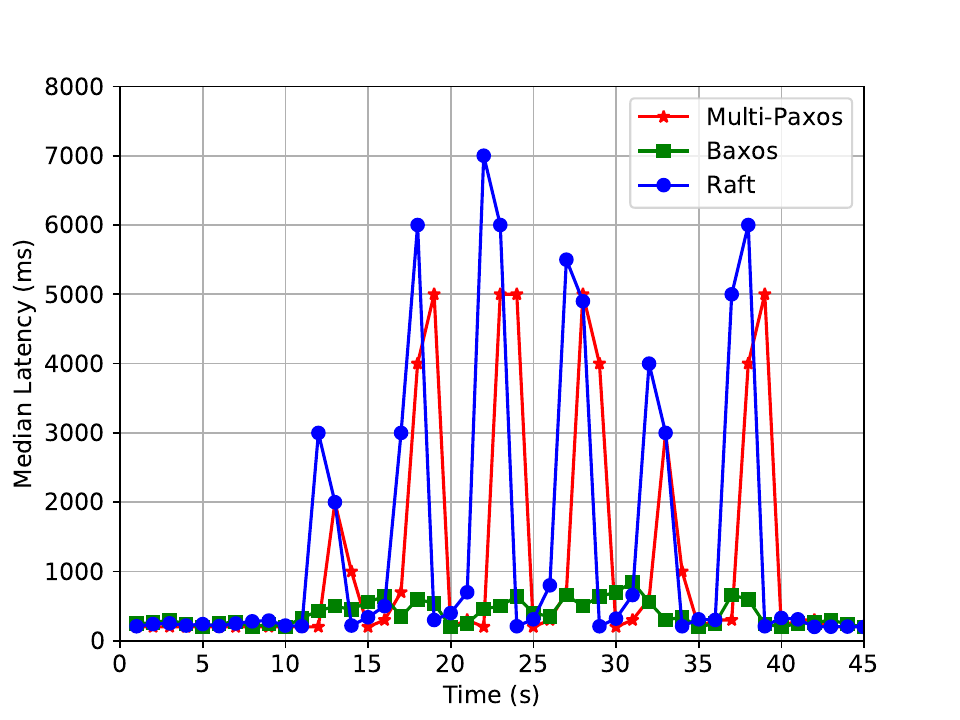}
      \caption{Median latency during delay attack}
      \vspace{-1mm}
    \end{subfigure} 
	\caption{Throughput and median latency under delay view change attack - Five clients in five AWS regions generate requests at 2,500 requests per second arrival rate (aggregate 12,500 requests per second on average). The attacker starts the attack at 10s, and keeps attacking the leader node in Multi-Paxos and Raft, and a random node in \sys for 30s. When the time is 40s, the attacker stops the attack. The workload consists of 8B request and response sizes with 1 $\mu$s service time. We use 5s view time out for Multi-Paxos and Raft in this experiment}
	\label{fig:attack_graph}
	\vspace{-5mm}

\end{figure*}

This experiment evaluates the performance of \sys under adversarial DDoS conditions. The attacker coordinates the attack by adaptively choosing the leader node and attacking it. In Multi-Paxos and Raft, the attacker targets the leader replica and dynamically adjusts the attack by following the current leader upon each view change. In \sys, there is no designated leader and the attacker chooses an arbitrary replica to attack. We experiment with two types of delayed view change attacks: (1) a \textit{delay attack}, where the adversary increases the transmission delay of a single replica to all destinations and (2) a \textit{packet loss attack}, where the adversary drops a fraction of egress packets of a single replica to all destinations. We used our micro benchmarks for this experiment. 

We observed the same throughput and median latency variation over time for both delay attack and the packet loss attack. Hence, we only show the delay attack results. ~\Cref{fig:attack_graph} compares the throughput and median latency of \sys under delay attack.

We first observe that during the first 10s of the experiment when there is no attack, all three consensus protocols progress at the speed of the network (the best case performance).  Second, we observe that the  throughput of Multi-Paxos, and Raft falls below  3,500 req/s on average, while Baxos delivers an average throughput of 8,000 requests per second, calculated over the attack time. Fourth, we observe that after 40s  (when the attack stops), all consensus algorithms eventually progress at the speed of the network.

We explain the throughput degradation of Multi-Paxos and Raft during the attack period as follows. In the delay attack, the attacker increases the latency of egress packets of the leader in Multi-Paxos and Raft up to 4s. In our experiments, we set the view timeout of Multi-Paxos and Raft to 5s. Since the maximum delay at the leader is less than the view change timeout, each replica receives some messages from the leader before a view change is triggered. To further avoid a view change, the attacker attacks the leader only up to 4s time period in a row, thus giving the leader node the opportunity to perform fast enough without being suspected by the follower nodes as a slow leader. Since the majority of the messages sent by the leader takes 4s on average, this reduces the speed of the entire replica set. This is the reason for observing a low throughput in Multi-Paxos and Raft, during the attack.

\sys achieves an \textit{average} throughput of 8,000 requests per second even in the presence of attacks. \sys does not employ a leader replica nor does it depend on the speed of all the nodes; \sys can make progress at the speed of the majority of replicas. Because the attacker attacks a single random replica at any given time, only the requests which are sent to the replica under attack experience high delays. The impact of the attack is negligible on the other replicas and clients. Hence Baxos delivers a throughput of 8,000 requests per second under attacks, on \textit{average}.

Figure~\ref{fig:attack_graph}(b) depicts the median latency of each consensus algorithm under study, with respect to time. We first observe that during attack-free executions, all consensus algorithms progress with a median latency less than 200ms. During the attack period (10s-40s), we observe that Multi-Paxos, and Raft deliver an median latency of 1250ms or higher, while Baxos has a median latency of 320ms. The reasoning for this behavior is same as the throughput discussion above: in Baxos the requests that are sent to the attacked replica experiences high delay whereas the requests sent to other replicas do not experience any high delay (thus low overall median latency). In Multi-Paxos and Raft, the latency of each request is affected by the attack given the leader-based message propagation.

\begin{figure}
  \includegraphics[width=\linewidth]{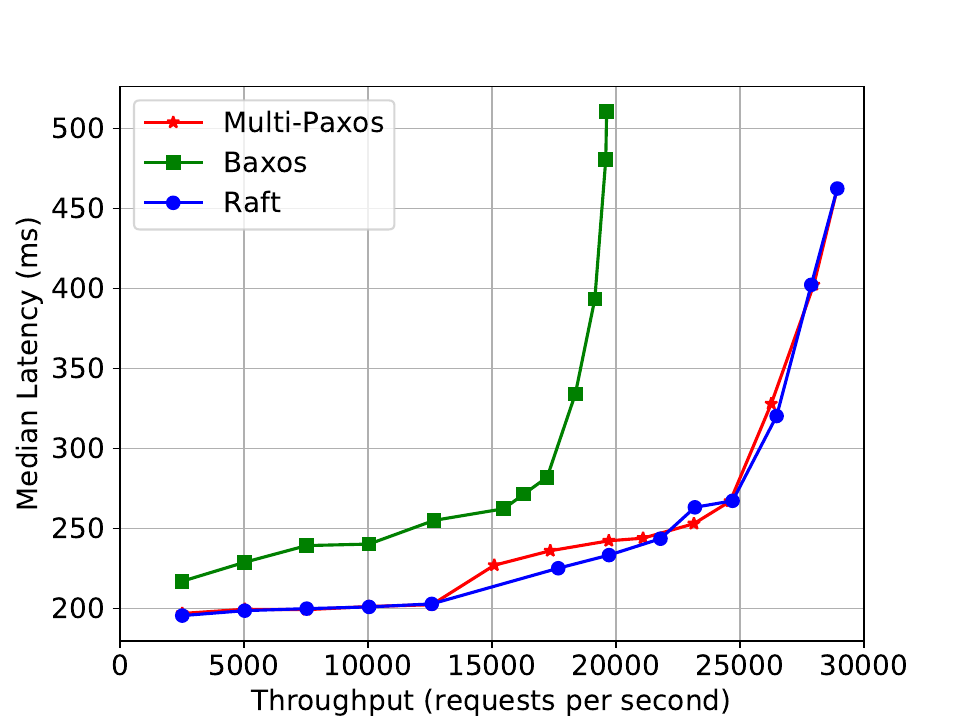}
  \caption{Normal-Case Performance. The workload consists of 8B request and response sizes with 1 $\mu$s service time. Five clients in 5 different AWS regions generate requests simultaneously -- exercises the worst case contention scenario (worst case performance).}
  \vspace{-5mm}
  \label{fig:best_case_performance}
\end{figure}

A variation of this attack is where the attacker crashes the leader node. To identify the impact of permanent leader failures, we conducted an experiment, where the leader replica is crashed. We use the same arrival rate and system parameters mentioned in Figure~\ref{fig:attack_graph} for this experiment. We observed that during the crash period (time period from the moment the leader is crashed and until a new leader is elected) Baxos delivers a steady throughput of 10,000 requests per second on \textit{average}, whereas Multi-Paxos and Raft cease to execute.

We conclude that \sys is up to 
128\% more resilient to DDoS attacks in throughput than Multi-Paxos and Raft.

\subsection{Attack-Free Case Performance} \label{sec:eval_sync}

This experiment aims at quantifying the performance overhead of \sys under faultless and synchronous network conditions. We use five client nodes that simultaneously send traffic to five replicas such that all five replicas propose commands. This experiment measures the worst case performance of \sys under highest possible contention. Since Baxos must resolve contention at each choice instead of relying on a stable leader, we expect Baxos to perform worse than leader-based algorithms under stable network conditions, but we wish to measure the performance cost of Baxos’s greater robustness. We used our micro benchmark for this experiment.

Figure~\ref{fig:best_case_performance} depicts the throughput vs. median latency graph. We observe that for a replica group of size five, \sys provides a maximum throughput of 17,500 requests per second under 300ms median latency, in contrast to 26,000 requests per second throughput of Multi-Paxos and Raft.

\paragraph{Throughput} The saturation throughput of \sys is 
32\% less than Multi-Paxos and Raft, because \sys faces contention: when multiple replicas propose requests simultaneously, their proposals collide, which leads to backing off by replicas and subsequent retries. While \sys's REB mechanisms enables us to reduce this contention, it cannot completely eliminate its impact. In contrast, Multi-Paxos and Raft do not experience contention because there is a single leader replica which proposes all commands.

\begin{table}
\centering
\begin{tabular}{|l|l|l|l|}
\hline
Baxos & Multi-Paxos & Raft   \\ \hline
354ms & 238ms       & 235ms   \\ \hline
\end{tabular}
\caption{Tail latency: The  workload  consists of  8B  request  and  response  sizes  with  1$\mu$s  service  time. Five  clients  in  5  different  AWS  regions  generate  requests at 2,500 requests per second (aggregate 12,500 requests per second) simultaneously}
\vspace{-5mm}
\label{tbl:tail}
\end{table}

\paragraph{Tail latency} Table~\ref{tbl:tail} illustrates the 99\% latency of each algorithm. We observe that the tail latency of Baxos is 
48\% higher than Multi-Paxos and Raft. 
The 48\% high tail latency of Baxos is caused by the re-transmissions: when Baxos faces contention it re-transmits, whereas in Multi-Paxos and Raft no request is re-transmitted in the best case execution.

These experimental findings show that \sys provides a low but acceptable performance to Multi-Paxos and Raft in the attack-free and synchronous network settings. 
We feel that this modest performance cost under high contention is justified in many applications, especially those in which load is sporadic and robustness under all conditions is important. In contrast, if the best case performance is the primary goal, Baxos is appealing as a fallback protocol under DDoS attacks: use Multi-Paxos under default synchronous network settings, and fall back to Baxos if there is a DDoS attack aimed at the leader.

\subsection{Bandwidth Utilization} \label{sec:effi}

\begin{figure}
  \vspace{-4mm}
  \includegraphics[width=0.8\linewidth]{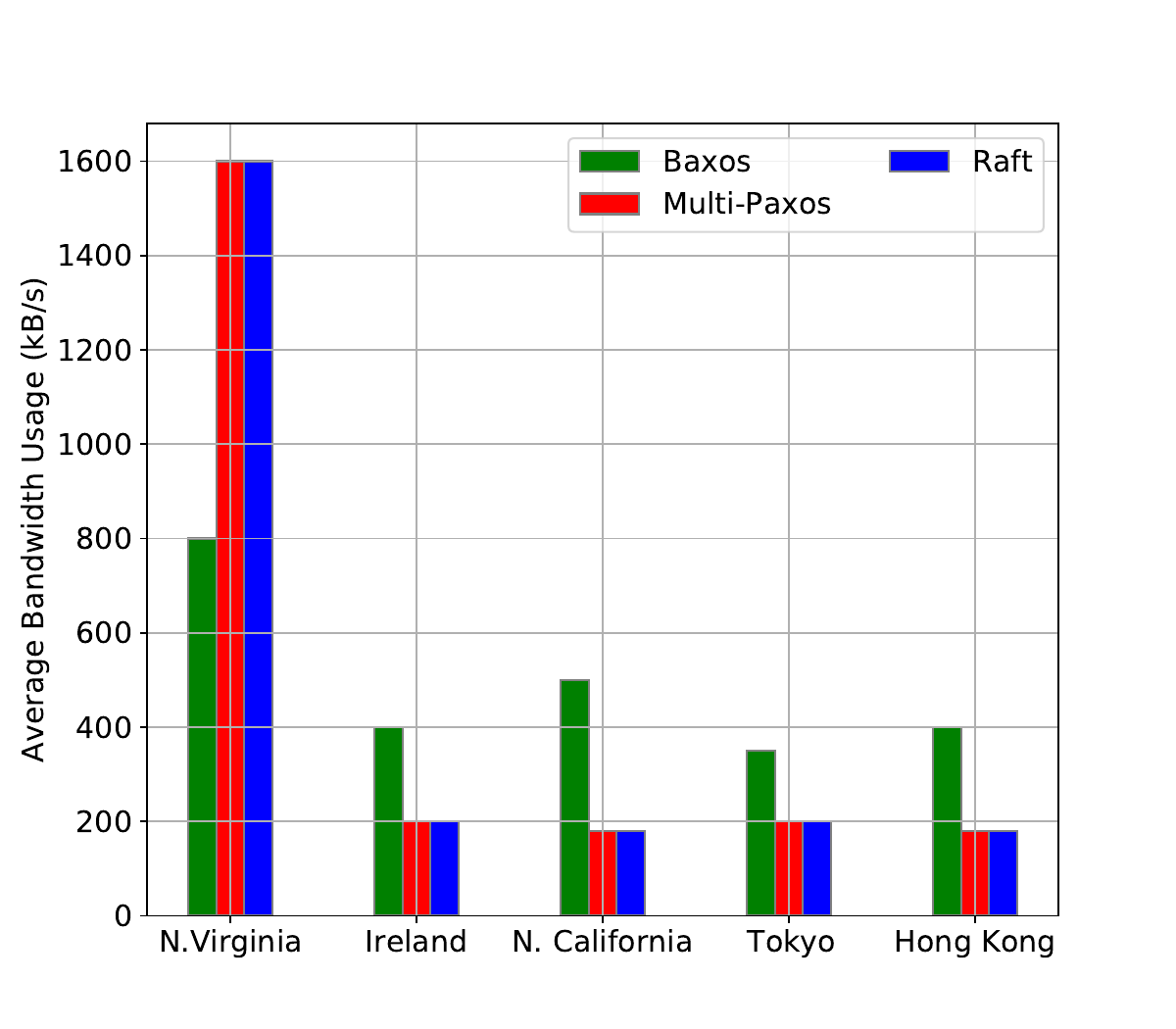}
  \vspace{-4mm}
  \caption{Average bandwidth usage of replicas. The leader nodes of Multi-Paxos and Raft algorithms are located in North Virginia. All five clients generate requests simultaneously at 1,000 requests per second}
  \label{fig:bandwidth}
  \vspace{-5mm}
\end{figure}

Efficiency of resource usage is an important but often overlooked aspect in consensus algorithms~\cite{matte2021scalable}. In addition to absolute measures of resources consumed, an efficient consensus algorithm should make each replica spend roughly the same amount of resources~\cite{barcelona2008mencius} resulting in a uniform resource usage. Uniform resource usage is important due to two main reasons: (1) a skew in resource usage results in a higher cost for power in data centers~\cite{prekas2015energy} and (2) in resource constrained setups, such as peer to peer systems where each node has the same amount of resources, it is prohibitive to have a high resource usage skew. To explore this property, we aim to answer the following question: \textit{What is the variability of resource usage of \sys replicas running in the wide-area?}

Since we experiment in the wide-area, where the performance is bottlenecked by the speed of the network, we only focus on the network I/O utilization.
We use our micro-benchmark for this experiment. To evaluate the variability of the  resource utilization by different replicas, we measure the ingress and egress traffic of each replica for a constant arrival rate. 

Figure~\ref{fig:bandwidth} depicts the bandwidth utilization of different replicas. For Multi-Paxos and Raft, the leader replica is located in North Virgina. We observe that Multi-Paxos and Raft consume 1,560 kB/s bandwidth on average in the leader replica while consuming less than 200 kB/s in non-leader replicas. 
In contrast, \sys consume 220-800 kB/s bandwidth in each replica, thus utilizing the bandwidth more uniformly across replicas. We explain these behaviors as follows.

In \sys, each replica proposes commands and on average, each replica sends and receives the same amount of messages per second. Hence, in \sys, each node roughly consumes the same amount of bandwidth. We calculated the standard deviation of the bandwidth utilization of different \sys replicas to be 152. In contrast, the leader replica in Multi-Paxos and Raft sends and receives more messages than other replicas. This causes Multi-Paxos and Raft to have a bandwidth standard deviation of 560, which is significantly higher than that of \sys. 

While Baxos nodes consume more bandwidth than non-leader nodes in Multi-Paxos and Raft, its utilization is relatively uniform across nodes and far lower than the leader’s bandwidth in leader-based schemes, which makes it a practical choice for data centers and resource constrained deployments such as sensor based internet of things applications.

\subsection{Scalability in Replica Set Size} \label{sec:scala}

This section evaluates the wide-area scalability of \sys using a standard benchmark. We evaluated the scalability of \sys, Multi-Paxos and Raft by running them with a replica set size of three, five, seven and nine. Unlike permissioned and permission-less blockchains where consensus algorithms are often designed to scale up to hundreds of nodes, crash fault tolerant protocols are usually designed to scale up to 9 nodes \cite{kogias2020hovercraft}\cite{barcelona2008mencius}. 
Hence, we evaluate Baxos only up to 9 nodes. We used nine AWS regions located in N. Virginia, Ireland, N. California, Tokyo, HongKong Oregon, Mumbai, Seoul and Cape Town.

We used Redis~\cite{carlson2013redis} with YCSB-A~\cite{cooper2010benchmarking} workload for this experiment. YCSB-A is a cloud benchmark workload that consists of a mix of 50/50 reads and writes modelling a session store recording recent actions. It assumes 1kB records with 10 fields of 100B each. The key selection is based on the Zipfian distribution. Redis is an in-memory key-value store that supports multiple data structures and operations, such as hash maps, sets and lists. We chose Redis as the backend application due to its wide adoption in the cloud performance analysis literature.

\begin{figure}
  \includegraphics[width=0.8\linewidth]{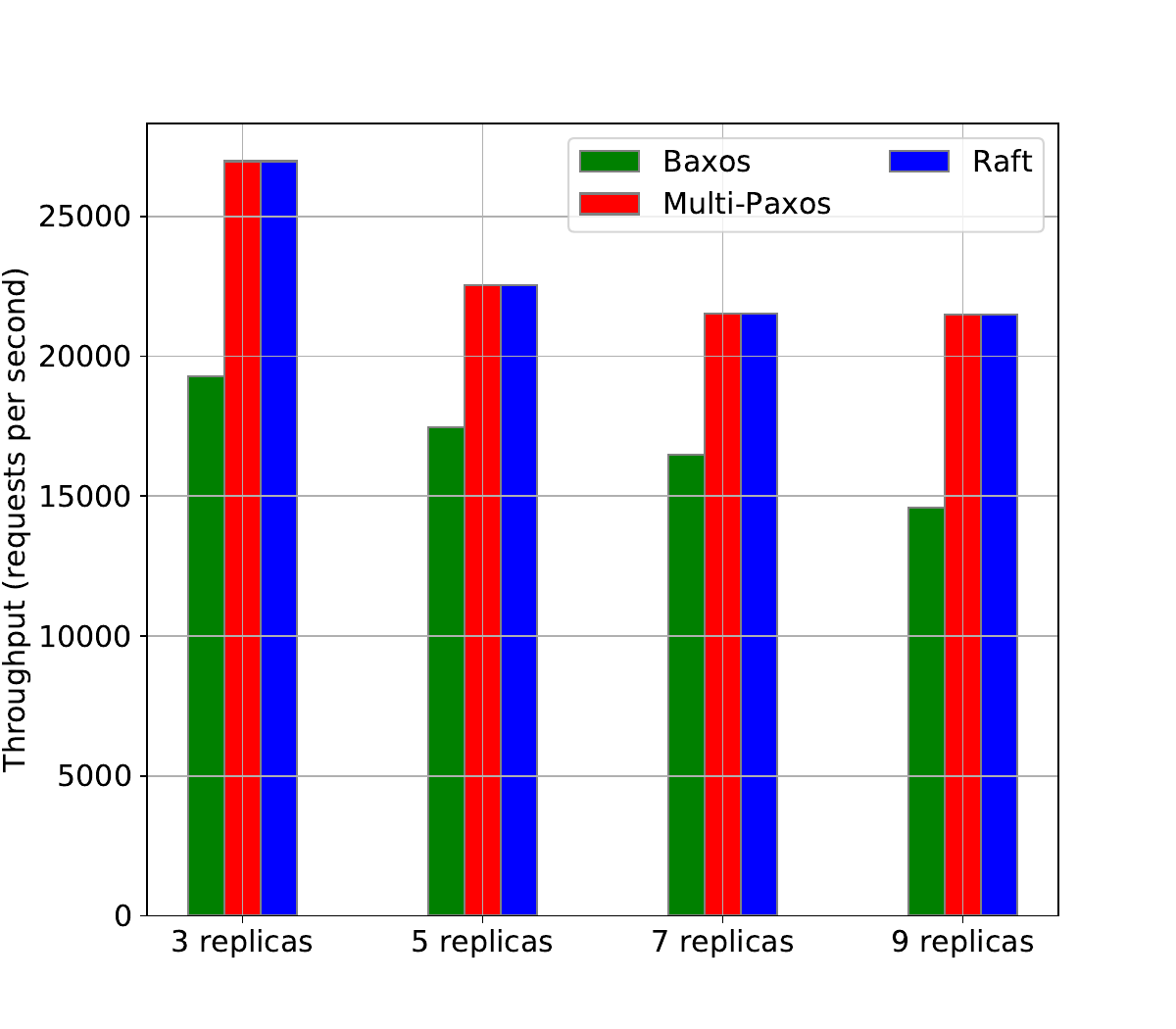}
  \caption{Scalability with respect to increasing replica count. The workload is YCSB-A workload with 1kB request size with Redis key value store as the backend. All throughput values are measured under 1 second 99\% percentile client perceived latency.}
  \label{fig:scalability}
\end{figure}

Figure~\ref{fig:scalability} depicts the scalability of \sys with respect to increasing replica counts in the wide-area. We first observe that the throughput values in this experiment are lower than the values presented in Figure~\ref{fig:best_case_performance}. Second, we observe that when the number of replicas is increased from three to nine, the throughput of each algorithm decreases; the throughput of \sys decreases from 19,000 to 14,500, whereas for Multi-Paxos and Raft the throughput decreases from 27,000 to 22,500) requests per second. 

We observe a reduction of the maximum throughput for all four consensus algorithms compared to the normal case performance experiment above (~\Cref{fig:best_case_performance}) due to the higher network bandwidth usage of this experiment. In the normal case performance experiment, we employed our micro benchmark with a 8B request size whereas in this scalability experiment we employed the YCSB-A workload with a 1kB request size.

The throughput of \sys, Multi-Paxos, and Raft decrease by 
21\%, 
20\%, and 20\% respectively, when the replica set size is increased from three to nine, due to two main reasons. First, with increasing replica count, the number of messages sent and received, when proposing a new command by the proposer, increases. Second, with an increasing replica count, the quorum size ($n/2 +1$) increases, thus the proposer has to collect Accept messages from a larger number of replicas. This affects the performance because in the wide-area experiments the proposer has to wait to collect responses from replicas located further away.

Using this empirical study on scalability, we conclude that \sys scales to a minimum of nine nodes while exhibiting the same percentage throughput loss with respect to the number of nodes as Multi-Paxos and 
Raft.
\section{Limitations and Future Work}\label{sec:limit}

We now discuss the limitations of \sys and the future work.

\textbf{Byzantine failures.} In this work, we only focus on crash failures. Despite our insights, it might not be straightforward to derive a Byzantine version of \sys because random backoff is not built on a quorum abstraction. Moreover, malicious parties can lie when they detect the contention and skew their ``start backoff time" as they please. 
We plan to explore Byzantine \sys using two approaches: (1) verifiable random functions~\cite{micali1999verifiable} and (2) trusted hardware base to enforce random backoff.

\textbf{Read Optimization.} In the current version of \sys, we do not differentiate between reads and writes, and both reads and writes are totally ordered using the same execution path. We intend to explore read optimizations using \textit{read leases}~\cite{moraru2013there}.

\textbf{Network Bandwidth Usage.} In \sys, Proposers have to broadcast a message to Acceptors, both in the Prepare-Promise phase and in the Propose-Accept phase, which results in a major I/O scalability bottleneck. We plan to address this issue in the future by exploring two approaches: (1) dynamic broadcast trees and (2) separating the total ordering from message broadcasting~\cite{danezis2021narwhal} by employing a peer to peer overlay to disseminate requests and using hash 
of requests for total ordering.
\section{Related Work}\label{sec:related}

\textbf{Liveness and performance downgrade attacks.} DDoS-resistant protocols based on a ``moving target''~\cite{nikolaou2015turtle,nikolaou2016moving} switch between different approaches depending on the network adversary. When the network is synchronous, these protocols employ single-decree Paxos, which delivers good performance in a synchronous network. When the system is under attack, they employ Ben-Or~\cite{ben1983another}, a randomized asynchronous consensus algorithm. While switching between these protocols provides a good performance when the network is synchronous, it performs poorly (but preserves liveness) when the network is experiencing transient but high delays because of the high message complexity of Ben-Or~\cite{ben1983another}. Moreover, this approach to DDoS resistance is challenging to implement due to complexities of merging two different consensus protocols. In contrast, \sys uses the same core consensus algorithm for the attack-free synchronous scenario and the DDoS attack scenario, resulting in fewer lines of code to implement and a better performance in the presence of transient high network delays. Spiegelman \textit{et al.}~\cite{spiegelman2019ace} have proposed a framework to transform a view based consensus protocol to a randomized consensus protocol to achieve robustness against DDoS attacks. However, their approach has a 100\% throughput overhead in the common case (synchronous) execution and as such, it is not suitable for applications requiring a good performance. In contrast, \sys has only a 7\% throughput overhead in the synchronous attack-free execution, compared to Multi-Paxos. Several other works, such as~\cite{amir2010prime}, have addressed the robustness of Byzantine consensus protocols under DDoS attacks but assuming a different threat model, where a Byzantine minority of replicas can misbehave. In \sys, we assumed that replicas are non-Byzantine.

\textbf{Use of REB and random timeouts in consensus algorithms.} Random exponential backoff and random timeouts have been explored in the context of consensus algorithms. IronFleet \cite{hawblitzel2015ironfleet} and PBFT \cite{castro1999practical} have employed random exponential timeouts to adapt the view change timeout with respect to the network conditions. This allows the replicas to adapt the timeout such that a quorum of Acceptors reply before a view change is triggered. Tendermint \cite{buchman2018latest} employs random timeouts inside a given consensus instance to prevent Tendermint from blocking forever for the liveness condition to be true, and to ensure that processes continuously transition between rounds. Renesse \textit{et al.} \cite{van2015paxos} use a simmilar approach to increase the time for which a leader waits to collect the responses from Acceptors. Renesse \textit{et al.} \cite{van2015paxos} employs a TCP-like additive increase, multiplicative decrease approach to select the optimal timeout to wait to collect reposnes from the Acceptors. Raft \cite{ongaro2014search} and Multi-Paxos \cite{lamport2001paxos} employs random timeouts to avoid concurrent and contending leader elections. Heterogeneous Paxos \cite{sheff2021heterogeneous} employs client side REB to avoid client induced flooding of the system. None of these approaches use REB as the primary method of contention handling, nor as a mechanism to withstand DDoS attacks. In contrast, Baxos employs REB as the primary method of contention handling to provide resielience agasint DoOS attacks.

\textbf{Leaderless consensus algorithms.} Mencius~\cite{barcelona2008mencius} achieves consensus without using a leader node by statically partitioning the log space among the set of replicas. This approach has two main drawbacks: (1) the speed of the system is dependent on the slowest replica and (2) an attack on a single replica can negatively affect the overall throughput of the system. In contrast, \sys makes progress at the speed of the majority of replicas, minimizing the effect of an attack on a single replica on the overall system. Generalized Paxos~\cite{lamport2005generalized} and  EPaxos~\cite{moraru2013there} achieve consensus without a leader by exploiting the request dependencies and using out-of-order commit. These protocols violate the layering constraints of a system design, which has lead to incorrect and complex specifications and implementations~\cite{efficientEpaxos,sutra2020correctness}. In contrast, \sys only needs to be aware of the total ordering of requests and it does not interfere with the application level dependencies such as the request commutativity. Fast Paxos~\cite{lamport2006fast} aims at reducing the number of round trips for a request from two to one but it fails to achieve a good performance in the presence of concurrent requests. Both Fast Paxos and Generalized Paxos assume a leader to resolve contention, hence, these protocols do not fully eliminate the leader bottleneck. Multi-coordinated Paxos~\cite{camargos2007multicoordinated} attempts to make the Generalized Paxos leaderless but it fails to deliver a good throughput as compared to Baxos, due to higher message complexity. 

\textbf{Other consensus variants.} Spaxos~\cite{biely2012s}, SDPaxos~\cite{zhao2018sdpaxos} and PigPaxos~\cite{charapko2021pigpaxos} aim at offloading the complexity of the leader node by separating the total ordering requirements from request propagation requirements; propagating a request among the replicas is done using a peer to peer overlay whereas the leader only proposes a total order for the digest of the request. However these approaches are vulnerable to delayed view change attacks because the total ordering is still dependent on a single leader replica. In contrast, \sys does not depend on a single replica for progression, achieving more robustness against DDoS attacks. Sharding has been studied in the context of consensus~\cite{ailijiang2017wpaxos, corbett2013spanner,  marandi2011high, dang2019partitioned, bezerra2014scalable, marandi2012multi} to improve the overall throughput by running multiple instances of the state machine. While sharding delivers high performance, it is an orthogonal problem to total ordering and employing sharding in the core consensus logic violates the layering argument. In contrast, sharding can be enabled in \sys by running \sys in each shard for DDoS resilience. Moreover, sharding based consensus algorithms are often arranged in a hierarchical fashion, where the lower level shards manage a shard of the total data set and the top level shard handles the inter-shard transactions. Sharding based protocols are vulnerable to DDoS attacks because an attack targeted at the top tier shard 
can bring down the performance of the entire system.
\section{Conclusion}\label{sec:conc}

We have presented Baxos, the first systematic exploration of the use of random exponential backoff (REB) in place of the usual leader election in practical consensus protocols. Our evaluation shows that \sys outperforms the commonly used leader-based consensus algorithms such as Multi-Paxos and Raft by 128\% in the presence of delayed view change attacks. 
We also explored the bandwidth efficiency of \sys and showed that \sys has a more uniform resource consumption than Raft and Multi-Paxos across replicas. Finally, we showed that \sys can scale up to nine replicas in the wide area.

\bibliographystyle{IEEEtran}

\end{document}